\documentclass[twocolumn,american,aps,prb,superscriptaddress]{revtex4-2}
\usepackage[T1]{fontenc}
\usepackage[utf8]{inputenc}
\usepackage{colortbl}
\usepackage{varwidth}
\usepackage{amsmath}
\usepackage{amssymb}
\usepackage{graphicx}

\makeatletter

\providecommand{\tabularnewline}{\\}
\newenvironment{cellvarwidth}[1][t]
    {\begin{varwidth}[#1]{\linewidth}}
    {\@finalstrut\@arstrutbox\end{varwidth}}

\usepackage{babel}

\makeatother

\usepackage{babel}
\begin{document}
\title{Role of Resonant $\mathbf{k}$-Points in the Transient Optical Response
of Pumped Germanium}
\author{Amir Eskandari-asl}
\affiliation{Dipartimento di Fisica ``E.R. Caianiello'', Università degli Studi
di Salerno, I-84084 Fisciano (SA), Italy}
\author{Giacomo Inzani}
\affiliation{Department of Physics and Regensburg Center for Ultrafast Nanoscopy
(RUN), University of Regensburg, D-93040 Regensburg, Germany}
\author{Matteo Lucchini}
\affiliation{Department of Physics, Politecnico di Milano, Piazza Leonardo da Vinci,
I-20133 Milano, Italy}
\affiliation{Institute for Photonics and Nanotechnologies, IFN-CNR, I-20133 Milan,
Italy}
\author{Adolfo Avella}
\affiliation{Dipartimento di Fisica ``E.R. Caianiello'', Università degli Studi
di Salerno, I-84084 Fisciano (SA), Italy}
\affiliation{CNR-SPIN, Unità di Salerno, I-84084 Fisciano (SA), Italy}
\affiliation{CNISM, Unità di Salerno, Università degli Studi di Salerno, I-84084
Fisciano (SA), Italy}
\begin{abstract}
Pump-induced transient optical properties combine contributions from
electronic states throughout the Brillouin zone, but the relative
relevance of off-resonant and $l$-photon resonant crystal momenta
has remained unexplored. We address this issue in pumped germanium
by resolving the transient absorptive response into momentum-space
classes defined by the presence or absence of 1-, 2-, and 3-photon
resonances with respect to the pump. Using the Dynamical Projective
Operatorial Approach together with the related generalized linear
response theory, we compute the differential imaginary part of the
dielectric function and evaluate the contributions of each resonance
class. Resonant regions account for nearly the entire optical response,
whereas points outside the identified resonance sets contribute only
negligibly. Nevertheless, the 2-photon-resonant set, although containing
more than 98\% of the residual (post-pump) excitation population,
does not reproduce the full transient spectrum. Conversely, resonance
classes with very small residual populations generate non-negligible
contributions to the transient optical properties. This mismatch shows
that the transient optical weight is not determined solely by the
real-charge dynamics (which results in post-pulse residual excitation
population) and is consistent with substantial virtual pump-induced
contributions, whose dominant optical weight nevertheless arises from
the resonant regions of momentum space. The class-resolved phase of
the dominant $2\omega_{\mathrm{pu}}$ oscillations further shows that,
whenever a class contributes appreciably, the phase of its oscillatory
component follows that of the corresponding full signal. The resulting
decomposition provides a momentum-resolved connection among multi-photon
resonances and transient optical observables in a realistic material.
\end{abstract}
\maketitle

\section{Introduction}

Ultrafast pump--probe spectroscopies provide access to the non-equilibrium
electronic response of solids on femtosecond and attosecond time scales
\cite{krausz2009attosecond,krausz2014attosecond,gandolfi2017emergent,geneaux2019,borrego2022attosecond,10.1063/5.0176656,Inzani_2025}.
By combining a strong driving field with a time-delayed probe, they
follow the evolution of electronic states during and after photo-excitation
and connect transient observables with microscopic charge, spin, and
lattice dynamics \cite{Zurch_17,doi:10.1063/1.4985056,PhysRevB.97.205202,perfetti2008femtosecond,schmitt2008transient}.
They also provide direct access to pump-induced symmetry breaking,
coherence, and relaxation processes \cite{Zurch_17,PhysRevB.97.205202,perfetti2008femtosecond}.

Transient optical spectroscopy and time-resolved angle-resolved photoemission
spectroscopy (TR-ARPES) offer complementary views of the non-equilibrium
dynamics. Optical measurements probe a response integrated over the
Brillouin zone \cite{schultze2013controlling,stojchevska2014ultrafast,schultze2014attosecond,lucchini2016attosecond,mashiko2016petahertz,Borja:16,zurch2017ultrafast,Zurch_17,schlaepfer2018attosecond,Kaplan:19,geneaux2019,inzani2023field,inzani2023photoinduced,neufeld2023attosecond,10.1063/4.0000253,Dolso:2025aa,inzani2026attosecond},
whereas TR-ARPES resolves transient occupations and dispersions in
momentum space \cite{schmitt2008transient,rohwer2011collapse,smallwood2012tracking,hellmann2012time,papalazarou2012coherent,wang2013observation,johannsen2013direct,rameau2016energy,reimann2018subcycle}.
Interpreting an optical signal therefore requires identifying which
regions of momentum space dominate each spectral structure. A second,
distinct question is whether that signal is only generated by real-carrier
dynamics, which in turn results in the residual carrier distribution
left after the pulse, or instead contains substantial contributions
from virtual processes that do not produce a comparable post-pulse
population. These issues are relevant to the use of intense light
fields as probes and control parameters in solids \cite{delatorre2021colloquium,Borsch:2023aa,Heide:2024aa}.
The underlying dynamics can involve intra-band motion \cite{schultze2013controlling,schlaepfer2018attosecond},
virtual carrier dynamics \cite{neufeld2023attosecond,Dolso:2025aa},
and resonant or non-resonant inter-band transitions \cite{schultze2013controlling,schlaepfer2018attosecond,inzani2023field,inzani2023photoinduced,inzani2026attosecond}.
Resolving their momentum-space contributions is thus necessary for
a microscopic interpretation of ultrafast optical spectra and light-driven
control \cite{krausz2014attosecond}, with possible implications for
ultrafast information technologies \cite{delatorre2021colloquium,Borsch:2023aa,Heide:2024aa}.

A quantitative understanding of these phenomena requires theoretical
approaches capable of describing the non-equilibrium electron dynamics
while retaining microscopic insight into the underlying processes.
Time-dependent density functional theory provides a first-principles
route \cite{de2013inside,de2016monitoring,wopperer2017efficient,de2017first,pemmaraju2018velocity,tancogne2018atomic,schlaepfer2018attosecond,neufeld2023attosecond,de2018real},
but in addition to its high computational cost, it is very difficult
for it to do a systematic decomposition of the response and the identification
of the processes responsible for individual spectral structures \cite{armstrong2021dialogue}.
Model-Hamiltonian approaches provide a complementary route, allowing
the role of individual microscopic processes to be investigated in
a controlled manner, although this often comes at the price of simplified
electronic structures \cite{armstrong2021dialogue}.

The Dynamical Projective Operatorial Approach (DPOA) was developed
to combine realistic electronic structures with microscopic resolution
\cite{inzani2023field,eskandari2024time,eskandari2024generalized,eskandari2024out,eskandari2025dynamical}.
It provides an operator-based description of the real-time evolution
of composite operators \cite{mancini2004} in multi-band systems driven
by intense fields while preserving direct access to microscopic observables.
Within this framework, one can calculate excitation populations \cite{inzani2023field,eskandari2024time,eskandari2025controlling,eskandari2026controlling},
TR-ARPES signals \cite{eskandari2024time}, and transient optical
responses \cite{eskandari2024generalized,eskandari2026magneto,inzani2026attosecond}
using a material-specific Hamiltonian. The formulation of DPOA is
sufficiently general to treat systems with arbitrary lattice structures,
numbers of electronic bands, and additional microscopic complexities.

Our previous study showed that DPOA reproduces the measured transient
optical response of photo-excited germanium \cite{inzani2026attosecond}.
The present work addresses a complementary question: what is the momentum-space
origin of the calculated response in terms of being or not in resonance
with the pump pulse? We classify each sampled $\mathbf{k}$-point
according to the presence or absence of different types of resonances
with the pump pulse, and use the resulting sets to decompose the transient
imaginary part of the dielectric function. This decomposition allows
us to identify which resonance classes dominate different spectral
features, how their contributions differ across the probe-energy range,
and, crucially, to what extent real and virtual excitations shape
the transient optical response. The comparison shows that momentum
regions carrying almost all residual carriers are not sufficient to
recover the full response, whereas classes with little post-pulse
population can still contribute significantly to the transient optical
response. The analysis therefore separates contributions tied directly
to the real charge excitation distribution from those naturally associated
with virtual pump-induced processes, whose dominant optical weight
nevertheless remains concentrated in resonant regions of momentum
space.

Section~\ref{sec:Theory} summarizes the DPOA description of the
driven system, the resonance-resolved residual populations, and the
generalized linear response used to calculate the transient optical
properties. Section~\ref{sec:Numerical-studies} introduces the electronic-structure
model, compares the calculated reflectivity with experiment, and presents
the momentum- and resonance-resolved analysis of the transient dielectric
response. Section~\ref{sec:Conclusions} summarizes the physical
conclusions. 

\section{Theory}\label{sec:Theory}

\subsection{Pump-Driven System}

We consider a crystalline lattice subjected to a linearly polarized
electromagnetic pulse, characterized by a vector potential $\boldsymbol{A}_{\text{pu}}\left(t\right)=A_{\text{pu}}\left(t\right)\hat{\boldsymbol{u}}_{\text{pu}}$
and an associated electric field $\boldsymbol{E}_{\text{pu}}\left(t\right)=E_{\text{pu}}\left(t\right)\hat{\boldsymbol{u}}_{\text{pu}}=-\partial_{t}A_{\text{pu}}\left(t\right)\hat{\boldsymbol{u}}_{\text{pu}}$.
The unit vector $\hat{\boldsymbol{u}}_{\text{pu}}$ specifies the
polarization axis of the pump pulse. The initial time $t_{\mathrm{ini}}$
is taken as $t_{\mathrm{ini}}\rightarrow-\infty$, when the system
is considered to be in equilibrium.

Within the dipole gauge, the time-dependent Hamiltonian governing
the system can be expressed as \cite{schuler2021gauge,eskandari2024time,eskandari2024generalized}
\begin{equation}
\mathcal{H}\left(t\right)=\sum_{\mathbf{k},n,n{}^{\prime}}c^{\dagger}_{\mathbf{k},n}\left(t\right)\Xi_{\mathbf{k},n,n{}^{\prime}}\left(t\right)c_{\mathbf{k},n{}^{\prime}}\left(t\right),
\end{equation}
where $c_{\mathbf{k},n}\left(t\right)$ destroys an electron with
crystal momentum $\mathbf{k}$ and composite index $n$. Throughout,
$n$ incorporates both band and spin degrees of freedom and is referred
to simply as the band index. The matrix $\Xi_{\mathbf{k}}\left(t\right)$
arises from the first-quantization single-particle Hamiltonian and
takes the form 
\begin{equation}
\Xi_{\mathbf{k}}\left(t\right)=T_{\mathbf{k}}\left(t\right)+\mathrm{e}E_{\text{pu}}\left(t\right)D_{\mathbf{k},\text{pu}}\left(t\right),
\end{equation}
with $\mathrm{e}>0$ the absolute electronic charge, $D_{\mathbf{k},\text{pu}}\left(t\right)=\boldsymbol{D}_{\mathbf{k}}\left(t\right)\cdot\hat{\boldsymbol{u}}_{\text{pu}}$.
Here, $T_{\mathbf{k}}\left(t\right)$ and $\boldsymbol{D}_{\mathbf{k}}\left(t\right)$
represent the time-dependent hopping and dipole matrices, respectively.
These quantities are obtained by first applying a generalized Peierls
substitution within a localized basis (typically the Wannier basis)
and subsequently transforming to the band basis, where the equilibrium
Hamiltonian becomes diagonal \cite{eskandari2024time,eskandari2024generalized}.

The unitary transformation $\Omega_{\mathbf{k}}$ relates the localized
(typically Wannier) basis to the band basis. In compact matrix notation,
$\Omega^{\dagger}_{\mathbf{k}}\cdot\tilde{T}_{\mathbf{k}}\cdot\Omega_{\mathbf{k}}$
is diagonal, with diagonal entries given by the band energies $\varepsilon_{\mathbf{k},n}$.
For an arbitrary matrix $M$ (which may represent $T$, $\boldsymbol{D}$,
or their derivatives with respect to $\mathbf{k}$), we define
\begin{equation}
M_{\mathbf{k}}\left(t\right)=\Omega^{\dagger}_{\mathbf{k}}\cdot\tilde{M}_{\mathbf{k}+\frac{\mathrm{e}}{\hbar}\boldsymbol{A}_{\text{pu}}\left(t\right)}\cdot\Omega_{\mathbf{k}}.
\end{equation}
Operators expressed in the localized basis are denoted by a tilde.
The symbol $\cdot$ denotes matrix multiplication in the electronic
Hilbert space and scalar contraction in Cartesian space. For realistic
multi-band systems, $\tilde{M}_{\mathbf{k}+\frac{\mathrm{e}}{\hbar}\boldsymbol{A}_{\text{pu}}\left(t\right)}$
is evaluated efficiently by expanding it in powers of the vector potential
through the so-called Peierls expansion \cite{eskandari2024time}.

Under the influence of the pump pulse, the DPOA allows us to express
the time-evolved operators $c_{\mathbf{k}}\left(t\right)$ in the
Heisenberg picture in terms of their equilibrium counterparts $c_{\mathbf{k}}\left(t_{\mathrm{ini}}\right)$
(where $c_{\mathbf{k}}\left(t\right)$ is a column vector with components
$c_{\mathbf{k},n}\left(t\right)$). This is accomplished via the projection
matrices $P_{\mathbf{k}}\left(t\right)$, which satisfy \cite{eskandari2024time}
\begin{equation}
c_{\mathbf{k}}\left(t\right)=P_{\mathbf{k}}\left(t\right)\cdot c_{\mathbf{k}}\left(t_{\mathrm{ini}}\right).
\end{equation}
The projection matrices obey the equation of motion 
\begin{equation}
\mathrm{i}\hbar\partial_{t}P_{\mathbf{k}}\left(t\right)=\Xi_{\mathbf{k}}\left(t\right)\cdot P_{\mathbf{k}}\left(t\right),
\end{equation}
which is numerically solved subject to the initial condition $P_{\mathbf{k}}\left(t_{\mathrm{ini}}\right)=\boldsymbol{1}$.

\subsection{Residual Photo-excited Population}

The electronic occupation of band $n$ at momentum $\mathbf{k}$ and
time $t$ is defined as $\mathcal{N}_{\mathbf{k},n}\left(t\right)=\langle c^{\dagger}_{\mathbf{k},n}\left(t\right)c_{\mathbf{k},n}\left(t\right)\rangle$.
In terms of the projection matrices, this quantity reads 
\begin{equation}
\mathcal{N}_{\mathbf{k},n}\left(t\right)=\sum_{n^{\prime}}P_{\mathbf{k},n,n^{\prime}}\left(t\right)f\left(\varepsilon_{\mathbf{k},n^{\prime}}\right)P^{\dagger}_{\mathbf{k},n^{\prime},n}\left(t\right),
\end{equation}
where $f\left(\varepsilon\right)=[e^{\beta\left(\varepsilon-\mu\right)}+1]^{-1}$
denotes the Fermi-Dirac distribution, with $\beta$ the inverse temperature
and $\mu$ the chemical potential.

We model the pump pulse with a Gaussian envelope: 
\begin{equation}
A_{\mathrm{pu}}\left(t\right)=A_{0}e^{-(4\ln2)t^{2}/\tau^{2}_{\mathrm{pu}}}\sin\left(\omega_{\mathrm{pu}}t\right),
\end{equation}
where $\tau_{\mathrm{pu}}$ is the full-width at half-maximum (FWHM)
duration of the pump pulse, and $\omega_{\mathrm{pu}}$ is its frequency.
The corresponding pump-pulse photon energy is $\hbar\omega_{\mathrm{pu}}$.

After the pulse has subsided, appreciable deviations of the residual
electronic population $\mathcal{N}^{\mathrm{res}}_{\mathbf{k},n}=\mathcal{N}_{\mathbf{k},n}\left(t\rightarrow\infty\right)$
from its equilibrium value $f\left(\varepsilon_{\mathbf{k},n}\right)$
are concentrated at momenta where inter-band energy differences satisfy
multi-photon resonance conditions with the pump \cite{inzani2023field,eskandari2024time,eskandari2025controlling}.
Here, $t\rightarrow\infty$ denotes a time after the pulse but before
relaxation mechanisms become relevant \cite{inzani2023field}. The
residual excitation population is defined as
\begin{equation}
N^{\mathrm{res}}_{\mathbf{k},n}=\mathcal{N}^{\mathrm{res}}_{\mathbf{k},n}-f\left(\varepsilon_{\mathbf{k},n}\right).
\end{equation}
Positive values $N^{\mathrm{res}}_{\mathbf{k},n}>0$ correspond to
photo-excited electrons in conduction bands (CBs), whereas negative
values $N^{\mathrm{res}}_{\mathbf{k},n}<0$ correspond to the associated
holes in valence bands (VBs). We focus on a gapped semiconductor at
low temperature, ensuring a clear separation between VBs and CBs.

The total residual excitation population per unit cell, $N^{\mathrm{res}}$,
is given by
\begin{equation}
N^{\mathrm{res}}=\frac{1}{\mathcal{M}_{\mathrm{grid}}}\sum_{\mathbf{k},n_{C}}N^{\mathrm{res}}_{\mathbf{k},n_{C}},
\end{equation}
where the sum over $n_{C}$ runs over all CBs and $\mathcal{M}_{\mathrm{grid}}$
denotes the total number of $\mathbf{k}$-points in the numerical
grid.

The weight of an $l$-photon resonance for a given gap energy $\varepsilon_{\mathrm{gap}}$
is expressed as \cite{eskandari2024time}
\begin{equation}
w_{l}(\varepsilon_{\mathrm{gap}})=\exp\left[-\frac{\tau^{2}_{\mathrm{pu}}}{8\ln2\hbar^{2}l}\left(\varepsilon_{\mathrm{gap}}-l\hbar\omega_{\mathrm{pu}}\right)^{2}\right].
\end{equation}
This form mirrors the squared spectral amplitude at frequency $\varepsilon_{\mathrm{gap}}/\hbar$
of the $l$-th power of the pump field, whose spectrum is centered
at $l\omega_{\mathrm{pu}}$.

The contribution associated with a transition from a specific VB,
$n_{V}$, to a specific CB, $n_{C}$, through an $l$-photon process
is estimated as \cite{inzani2023field,eskandari2024time,eskandari2026controlling}
\begin{equation}
N^{\mathrm{res}(l)}_{\mathbf{k},n_{C},n_{V}}=\frac{N^{\mathrm{res}}_{\mathbf{k},n_{V}}w_{l}(\varepsilon_{\mathbf{k},n_{C}}-\varepsilon_{\mathbf{k},n_{V}})}{\sum_{n_{V}'}N^{\mathrm{res}}_{\mathbf{k},n_{V}'}\sum_{l'}w_{l'}(\varepsilon_{\mathbf{k},n_{C}}-\varepsilon_{\mathbf{k},n_{V}'})}N^{\mathrm{res}}_{\mathbf{k},n_{C}}.
\end{equation}
Summing over all VBs yields the total $l$-photon contribution to
the residual excitation in $n_{C}$: 
\begin{equation}
N^{\mathrm{res}(l)}_{\mathbf{k},n_{C}}=\sum_{n_{V}}N^{\mathrm{res}(l)}_{\mathbf{k},n_{C},n_{V}}.
\end{equation}
Summing further over all CBs gives the total $l$-photon contribution
to the residual excitation at momentum $\mathbf{k}$: 
\begin{equation}
N^{\mathrm{res}(l)}_{\mathbf{k}}=\sum_{n_{C}}N^{\mathrm{res}(l)}_{\mathbf{k},n_{C}}.
\end{equation}

The corresponding $l$-photon contribution to the residual excitation
population per unit cell is 
\begin{equation}
N^{\mathrm{res}(l)}=\frac{1}{\mathcal{M}_{\mathrm{grid}}}\sum_{\mathbf{k}}N^{\mathrm{res}(l)}_{\mathbf{k}}.
\end{equation}

\subsection{Transient Optical Response}

We restrict the analysis to the collinear configuration in which the
pump, incoming-probe, and detected-probe polarization directions are
all the same, and denote this common direction by the unit vector
$\hat{\boldsymbol{u}}_{\text{pu}}$. Consequently, for the optical
conductivity tensor $\boldsymbol{\sigma}$, the relevant component
is $\sigma=\hat{\boldsymbol{u}}_{\text{pu}}\cdot\boldsymbol{\sigma}\cdot\hat{\boldsymbol{u}}_{\text{pu}}$.

Within the framework of the generalized linear response theory \cite{eskandari2024generalized}
and employing the DPOA, the non-equilibrium optical conductivity of
the pumped system at probe time $t_{\mathrm{pr}}$ (defined as the
center of the probe pulse), denoted $\sigma\left(\omega,t_{\mathrm{pr}}\right)$,
can be written as \cite{eskandari2024generalized}
\begin{equation}
\sigma\left(\omega,t_{\mathrm{pr}}\right)=\frac{1}{\mathcal{M}_{\mathrm{grid}}}\sum_{\mathbf{k}}\sigma_{\mathbf{k}}\left(\omega,t_{\mathrm{pr}}\right),
\end{equation}
where $\omega$ is the probe frequency, and $t_{\mathrm{pr}}$ corresponds
to the probe delay time in our convention where the pump pulse center
is set as the time origin. The quantity $\sigma_{\mathbf{k}}\left(\omega,t_{\mathrm{pr}}\right)$
represents the contribution from crystal momentum $\mathbf{k}$ to
the total optical conductivity. The complete procedure for computing
$\sigma_{\mathbf{k}}\left(\omega,t_{\mathrm{pr}}\right)$ using $P_{\mathbf{k}}\left(t\right)$
and generalized linear response theory is detailed in Refs.~\cite{eskandari2024generalized,eskandari2025dynamical,eskandari2026magneto}.

The non-equilibrium dielectric function is obtained from the optical
conductivity: 
\begin{equation}
\epsilon\left(\omega,t_{\mathrm{pr}}\right)=1+\frac{\mathrm{i}}{\omega\epsilon_{0}}\sigma\left(\omega,t_{\mathrm{pr}}\right),
\end{equation}
and subsequently, the reflectivity at an incident probe angle $\theta$,
$R_{\theta}\left(\omega,t_{\mathrm{pr}}\right)$ is computed from
$\epsilon\left(\omega,t_{\mathrm{pr}}\right)$ \cite{eskandari2024generalized,eskandari2025dynamical}.
The relative differential reflectivity is defined as 
\begin{equation}
\delta_{r}R_{\theta}\left(\omega,t_{\mathrm{pr}}\right)=\left[R_{\theta}\left(\omega,t_{\mathrm{pr}}\right)-R^{\mathrm{eq}}_{\theta}\left(\omega\right)\right]/R^{\mathrm{eq}}_{\theta}\left(\omega\right),
\end{equation}
where the superscript ``eq'' denotes equilibrium quantities. Of
particular interest in this work is the differential imaginary part
of the dielectric function: 
\begin{equation}
\delta\Im\epsilon\left(\omega,t_{\mathrm{pr}}\right)=\Im\epsilon\left(\omega,t_{\mathrm{pr}}\right)-\Im\epsilon^{\mathrm{eq}}\left(\omega\right).
\end{equation}
This quantity characterizes the absorptive component of the pump-induced
optical response \cite{eskandari2024generalized,eskandari2025dynamical}
and can be written as 
\begin{equation}
\delta\Im\epsilon\left(\omega,t_{\mathrm{pr}}\right)=\frac{1}{\mathcal{M}_{\mathrm{grid}}}\frac{1}{\omega\epsilon_{0}}\sum_{\mathbf{k}}\Re\left[\sigma_{\mathbf{k}}\left(\omega,t_{\mathrm{pr}}\right)-\sigma^{\mathrm{eq}}_{\mathbf{k}}\left(\omega\right)\right].\label{eq:d_Im_eps_k}
\end{equation}

The summation in Eq.~\eqref{eq:d_Im_eps_k} runs over the entire
$\mathbf{k}$-grid. In the following, we also evaluate partial sums
restricted to selected resonance classes in order to isolate their
individual contributions to the transient optical response $\delta\Im\epsilon$.
The normalization factor $\frac{1}{\mathcal{M}_{\mathrm{grid}}}$
is kept unchanged so that the contributions from complementary resonance
classes add to the full response.

\section{Resonance-Class Decomposition of the Optical Response of Pumped Germanium
}\label{sec:Numerical-studies}

We now apply the theoretical framework described in Sec.~\ref{sec:Theory}
to pumped germanium. The electronic structure is obtained from first-principles
calculations with the Elk code \cite{elk_code}, including spin-orbit
coupling and 36 spin-resolved bands: 20 $3d$ semi-core bands, 8 valence
and 8 conduction $sp^{3}$ bands \cite{inzani2023field,inzani2026attosecond,eskandari2026controlling}.

The hopping parameters, $\tilde{T}_{\mathbf{k}}$, and dipole matrix
elements, $\tilde{\mathbf{D}}_{\mathbf{k}}$, are obtained from a
Wannier representation constructed with Wannier90 \cite{pizzi2020wannier90},
following the procedure detailed in Refs.~\cite{inzani2023field,inzani2026attosecond,eskandari2026controlling}.
The Brillouin zone (BZ) is sampled on a $32\times32\times32$ $\mathbf{k}$-grid
centered at $\Gamma$, and all calculated bands, including the semi-core
manifold, are retained in the dynamics. A damping factor of $0^{+}=0.25$
PHz is used in the Fourier transformation of the optical conductivity.
Because the ab-initio calculations place the relevant semi-core manifold
at an inaccurate binding energy, the probe-energy axis is shifted
by $+5.3$ eV so that the calculated absorption edge coincides with
the experimental one. The crystal orientation and pump-pulse polarization
are chosen as in Refs.~\cite{inzani2023field,inzani2026attosecond,eskandari2026controlling},
with the field polarized along the {[}100{]} direction. The pump vector
potential has a FWHM duration $\tau_{\mathrm{pu}}=13.3$ fs, amplitude
$A_{0}=0.528$ V\,fs/nm, and photon energy $\hbar\omega_{\mathrm{pu}}=1.55$
eV.

\subsection{Equilibrium Band Structure}

\begin{figure}
\centering{}\includegraphics[width=8cm]{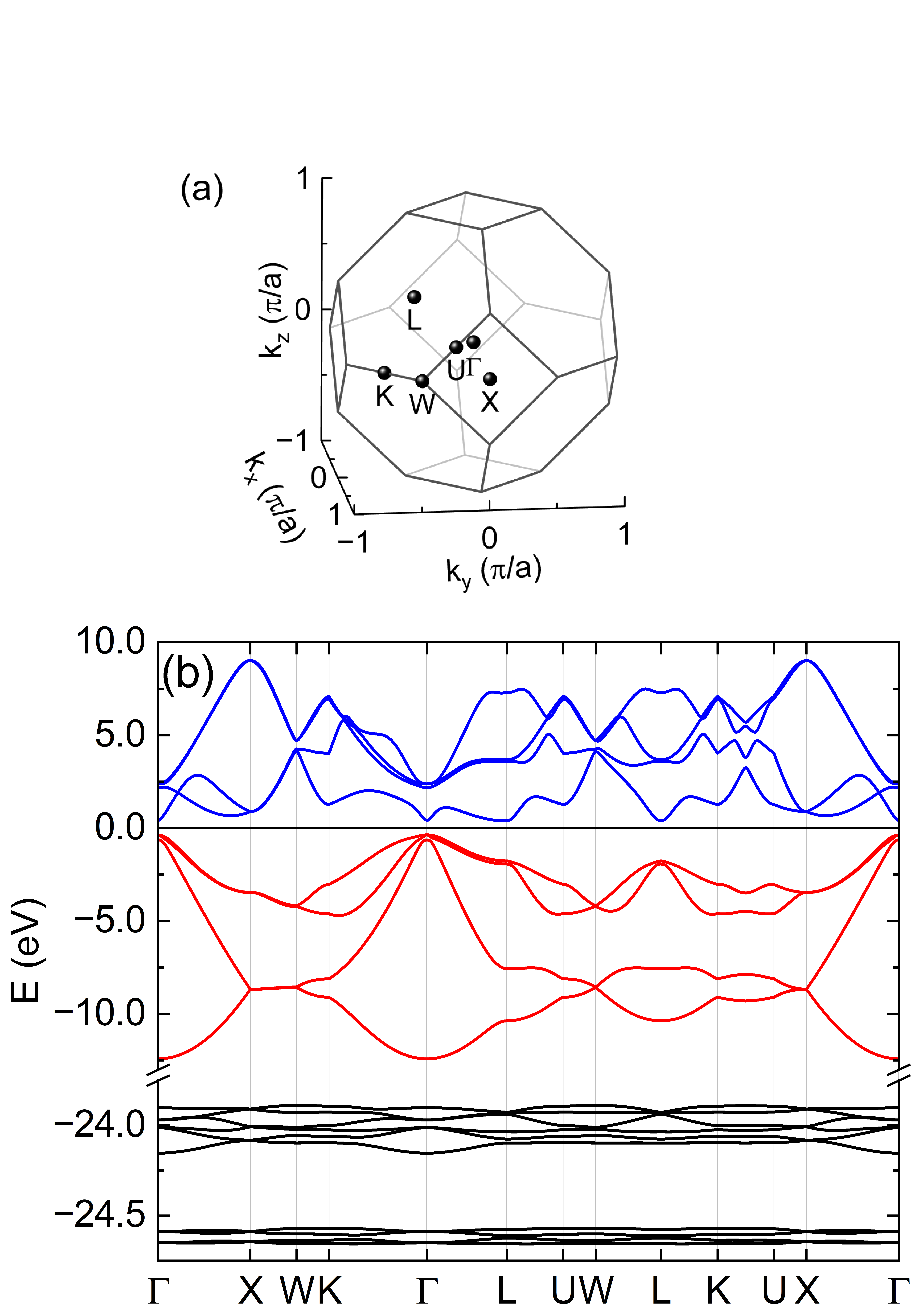}\caption{(a) First Brillouin zone of the fcc lattice, with the high-symmetry
points. (b) Equilibrium band structure of germanium along the indicated
high-symmetry path. The calculation contains the valence, conduction,
and semi-core manifolds relevant to the pump and probe transitions.
The $\Gamma_{2}$ gap is underestimated by approximately $0.5$ eV,
and the semi-core manifold is displaced relative to experiment. }\label{fig:FBZ_eqbands}
\end{figure}

Before analyzing the transient optical response, we briefly summarize
the equilibrium electronic structure of germanium, which provides
the basis for the subsequent momentum-resolved analysis (Fig.~\ref{fig:FBZ_eqbands}).
Figure~\ref{fig:FBZ_eqbands}(a) shows the first Brillouin zone (FBZ)
of the face-centered cubic (fcc) lattice, with the high-symmetry points
indicated. These define the reciprocal-space path along which the
equilibrium band structure is presented in Fig.~\ref{fig:FBZ_eqbands}(b).
Several features of the equilibrium band structure are particularly
relevant for the subsequent analysis. First, the $\Gamma_{2}$ gap
(the gap at $\Gamma$ between the uppermost VB and the second-lowest
CB) obtained from the ab initio calculations is smaller than the experimentally
reported value \cite{Ioffe-Germanium} by approximately 0.5 eV. Second,
as previously mentioned, the ab initio calculations place the semi-core
manifold at an inaccurate binding energy; this manifests as a shift
in the semi-core bands that requires a correction of $+5.3$ eV in
the probe energy axis to match the experimental absorption edge.

\begin{figure*}
\centering{}\includegraphics[width=12cm]{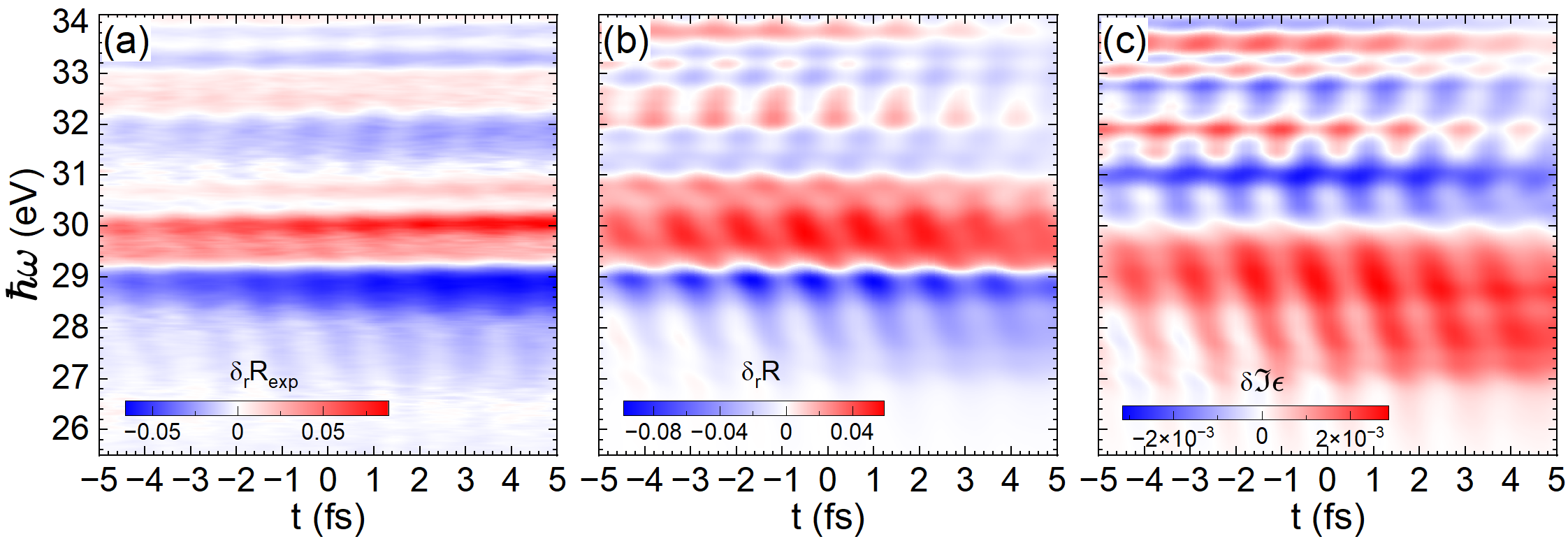}\caption{Experimental (a) and calculated (b) non-equilibrium relative differential
reflectivity. (c) Calculated differential imaginary part of the dielectric
function. All quantities are shown as functions of probe photon energy
$\hbar\omega$ and delay time $t_{\mathrm{pr}}$. The calculations
in panel (b) reproduce the principal experimental structures and their
temporal evolution; the high-energy structures retain an approximately
$0.5$ eV offset associated with the underestimated ab-initio $\Gamma_{2}$
gap. }\label{fig:exp_theo_refl}
\end{figure*}

\subsection{Transient Optical Response and Benchmarking}

Before resolving the response in momentum space, we first verify that
the same electronic-structure model reproduces the measured optical
signal. Figure~\ref{fig:exp_theo_refl}(a) shows the experimental
non-equilibrium relative differential reflectivity, with experimental
details reported in Refs.~\cite{inzani2023field,inzani2026attosecond}.
Panel (b) gives the DPOA result obtained from generalized linear response
theory, and panel (c) shows the corresponding absorptive quantity
$\delta\Im\epsilon(\omega,t_{\mathrm{pr}})$.

The calculation reproduces the principal experimental structures and
their evolution over the measured delay range \cite{inzani2026attosecond}.
Above approximately $32$ eV, the calculated structures have a downward
energy mismatch of about $0.5$ eV. This offset follows from the underestimated
ab-initio $\Gamma_{2}$ gap and therefore from the reduced energy
separation between the semi-core manifold and the relevant CBs.

Unlike the differential reflectivity, the differential imaginary part
of the dielectric function allows for a more direct interpretation.
Regions with negative $\delta\Im\epsilon$ correspond to a reduction
of the absorptive response. In the present energy range, these structures
are consistent with pump-induced state filling: carriers promoted
to the conduction bands reduce the number of available final states
for subsequent semi-core-to-CB probe transitions. The energy $\hbar\omega\simeq29.8$
eV marks an approximate crossover: above it, the dominant probe transitions
terminate in CB states, whereas below it they predominantly terminate
in VB states.

At the chosen incidence angle, the reflectivity depends predominantly
on the real part of the complex dielectric function \cite{10.1063/5.0176656},
which is related to the imaginary part by causality through the Kramers--Kronig
relations. Consequently, as it can be observed by comparing panels
(b) and (c), the extrema in each map coincide with the zero-crossing
contours in the other, consistently with both quantities being derived
from the same causal dielectric response \cite{eskandari2024generalized}.
The very good agreement between the theoretical relative differential
reflectivity and the experimental data provides an important validation
of the present approach. Since $\delta\Im\epsilon$ and the reflectivity
are linked through the Kramers--Kronig relations, this agreement
provides a consistent foundation for the momentum-resolved analysis
presented in the following sections.

\begin{figure*}
\centering{}\includegraphics[width=18cm]{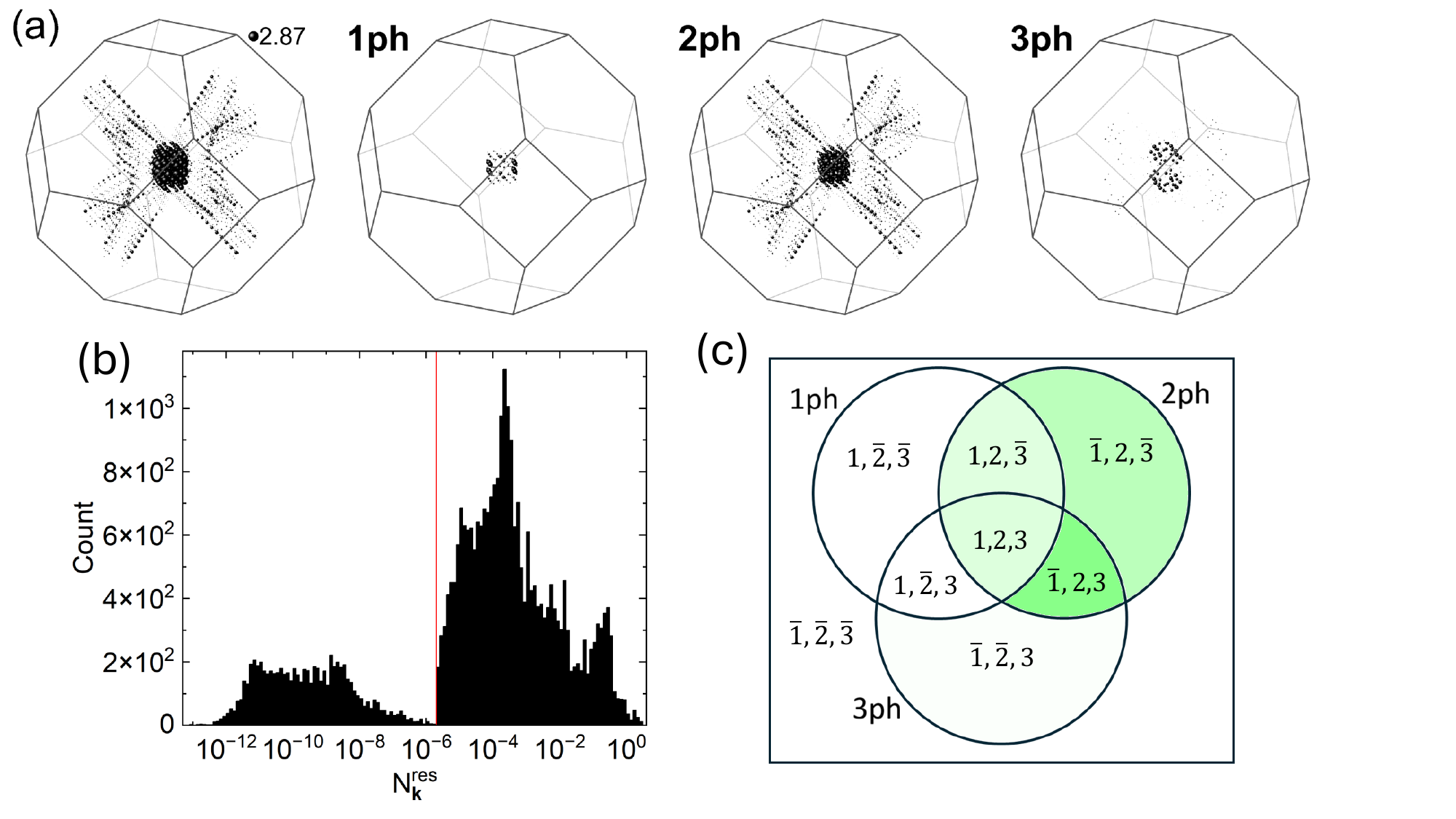}\caption{(a) Momentum-resolved residual excitation population across the FBZ
and its decomposition into 1-, 2-, and 3-photon contributions shown
as a 3D scatter plot. The diameter of the spheres indicate the actual
residual excitation population at the corresponding $\mathbf{k}$-point
and scales as it is reported in the legend. (b) Histogram of residual
excitation values on a logarithmic scale; the vertical red line at
$2\times10^{-6}$ marks the numerical threshold used to identify resonant
$\mathbf{k}$-points. (c) Venn diagram showing the relative contributions
of different resonance classes to the total residual excitation population.
The green shading represents the relative contribution, with quantitative
values given in Table~\ref{tab:Number-of--points}. }\label{fig:res_K}
\end{figure*}

\begin{table}[t]
\centering{}%
\begin{tabular}{ccc}
\hline 
\begin{cellvarwidth}[t]
\centering
 Resonance

class 
\end{cellvarwidth} & \begin{cellvarwidth}[t]
\centering
 Number of

$\mathbf{k}$-points

(tot. 32768) 
\end{cellvarwidth} & \begin{cellvarwidth}[t]
\centering
 Fraction of total

residual population 
\end{cellvarwidth}\tabularnewline
\hline 
$(1,2,3)$  & 220  & 0.13281\tabularnewline
$(1,2,\bar{3})$  & 64  & 0.12001\tabularnewline
$(1,\bar{2},3)$  & 16  & 0.00041\tabularnewline
$(1,\bar{2},\bar{3})$  & 0  & 0\tabularnewline
$(\bar{1},2,3)$  & 4830  & 0.46373\tabularnewline
$(\bar{1},2,\bar{3})$  & 1859  & 0.26572\tabularnewline
$(\bar{1},\bar{2},3)$  & 13445  & 0.01707\tabularnewline
$(\bar{1},\bar{2},\bar{3})$  & 12334  & 0.00024\tabularnewline
\hline 
\end{tabular}\caption{Number of $\mathbf{k}$-points and relative contribution to the residual
excitation population per unit cell, $N^{\mathrm{res}}$, for each
resonance class. An overbar denotes the absence of the corresponding
resonance type; for instance, $(1,2,\bar{3})$ indicates a class of
$\mathbf{k}$-points where 1- and 2-photon resonances are present
but 3-photon resonances are absent.}\label{tab:Number-of--points}
\end{table}

\subsection{Resonance Analysis and Residual Excitations}

Figure~\ref{fig:res_K} summarizes the momentum-resolved residual
excitation population together with its decomposition into multi-photon
resonance contributions. Panel (a) shows $N^{\mathrm{res}}_{\mathbf{k}}$
and its 1-, 2-, and 3-photon components throughout the sampled BZ.
As established in our previous work \cite{inzani2023field,eskandari2026controlling},
the dominant contribution arises from two-photon transitions.

Panel (b) displays the distribution of $N^{\mathrm{res}(l)}_{\mathbf{k}}$
on a logarithmic scale. We use $2\times10^{-6}$, marked by the red
vertical line, as the numerical threshold for assigning a $\mathbf{k}$-point
to the corresponding $l$-photon-resonant set. Throughout this work,
we adopt this threshold as the operational criterion for identifying
resonant momentum points: a point belongs to the $l$-photon set when
its decomposed contribution $N^{\mathrm{res}(l)}_{\mathbf{k}}$ exceeds
this value, and it may belong to more than one resonance set. Our
numerical precision for computing the residual excitations is about
$\sim10^{-7}$.

Panel (c) shows the mutually exclusive intersections of the 1-, 2-,
and 3-photon sets. The shading represents the fraction of the total
residual population per unit cell, associated with each class, and
the corresponding values are listed in Table~\ref{tab:Number-of--points}.
The four classes containing a 2-photon resonance together account
for more than 98\% of $N^{\mathrm{res}}$. By contrast, the other
classes contribute little to the post-pulse population, despite occupying
sizable portions of the sampled momentum space.

\begin{figure*}
\centering{}\includegraphics[width=16cm]{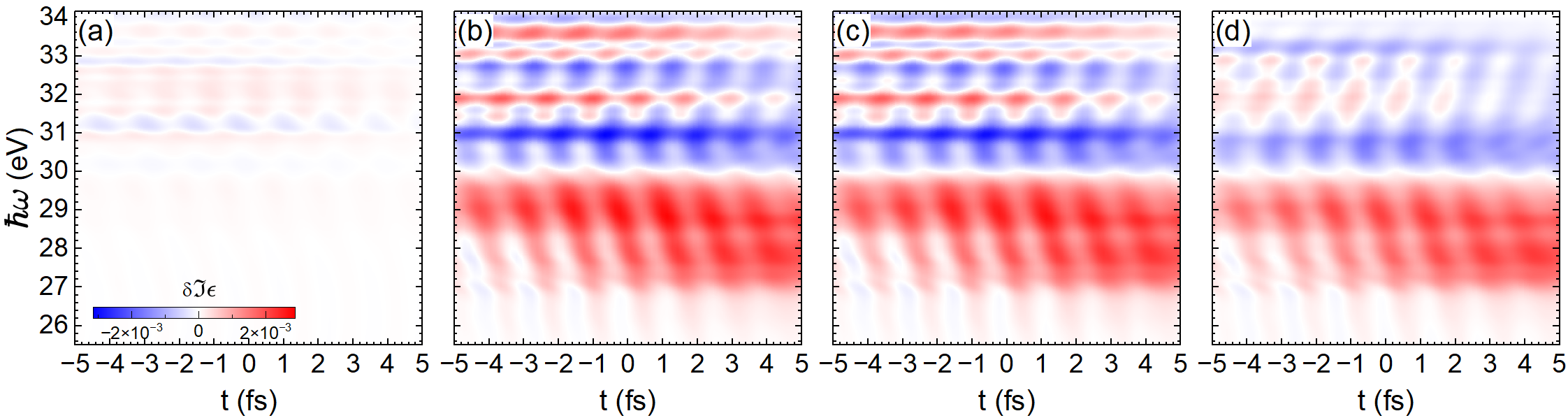}\caption{Contributions to $\delta\Im\epsilon(\omega,t_{\mathrm{pr}})$ from
different classes of $\mathbf{k}$-points: (a) off-resonant $\mathbf{k}$-points
only; (b) resonant $\mathbf{k}$-points; (c) $\mathbf{k}$-points
with 1-, 2-, or 3-photon resonances; (d) $\mathbf{k}$-points with
2-photon resonances. The resonant $\mathbf{k}$-points nearly reproduce
the full response {[}Fig.~\ref{fig:exp_theo_refl}(c){]}, whereas
the 2-photon-resonant set, despite carrying most of the residual population,
does not account for the complete response. }\label{fig:Optics_res_k}
\end{figure*}

\subsection{Role of Resonant and Off-Resonant $\mathbf{k}$-Points in the Transient
Optical Response}

Having established the resonance classification, we now use it to
disentangle the momentum-space origin of the transient optical response.
Specifically, we evaluate $\delta\Im\epsilon(\omega,t_{\mathrm{pr}})$
by restricting the summation in Eq.~\eqref{eq:d_Im_eps_k} to selected
momentum-space classes, thereby isolating their individual contributions.

Figure~\ref{fig:Optics_res_k}(a) shows the contribution from the
off-resonant $\mathbf{k}$-points included in the analysis. It is
small but nonzero. Panel (b) contains all points classified as resonant,
including higher-order processes, and nearly reproduces the complete
response {[}Fig.~\ref{fig:exp_theo_refl}(c){]}. The comparison with
the complete response shows that the transient absorptive signal is
mostly originated from the resonant regions of momentum space. In
particular, the $\Gamma$ point is off-resonant and makes no visible
contribution to the structures considered here.

Panel (c) retains the union of the 1-, 2-, and 3-photon sets, corresponding
to the area enclosed by the three circles in Fig.~\ref{fig:res_K}(c).
Its close agreement with the full response indicates that these lower-order
sets contain the dominant contributions in the investigated probe-energy
window. Panel (d), by contrast, retains only points belonging to the
2-photon set {[}i.e., $2$ph circle in Fig.~\ref{fig:res_K}(c){]}.
Although these points carry more than 98\% of the residual population
{[}Fig.~\ref{fig:res_K}(c) and Table~\ref{tab:Number-of--points}{]},
they do not reproduce the complete optical response.

This comparison demonstrates that the real-charge dynamics, which
results in residual excitation population, although essential, is
not sufficient to account for the complete transient optical response.
Any physical interpretation based solely on the real-charge dynamics
therefore misses a substantial part of the optical signal.

This observation has an important physical implication. The dominance
of resonant momentum-space regions in the transient optical response
should not be interpreted as evidence that the response is governed
primarily by the real photo-excited carriers. Instead, the comparison
between panels (c) and (d) shows that a substantial fraction of the
response is naturally attributed to virtual pump-induced processes
also in semiconductors and not only in insulators. Importantly, these
additional contributions still arise predominantly from the same resonant
regions of momentum space rather than from off-resonant ones.

This observation provides the physical basis for the more detailed
resonance-class decomposition presented below, whose purpose is to
identify how the different resonant momentum-space regions contribute
to the real and virtual components of the transient optical response.

Consequently, the resonance-class decomposition provides a unified
momentum-space framework for interpreting both the real and virtual
pump-induced contributions to the transient optical response.

\begin{figure*}
\centering{}\includegraphics[width=12cm]{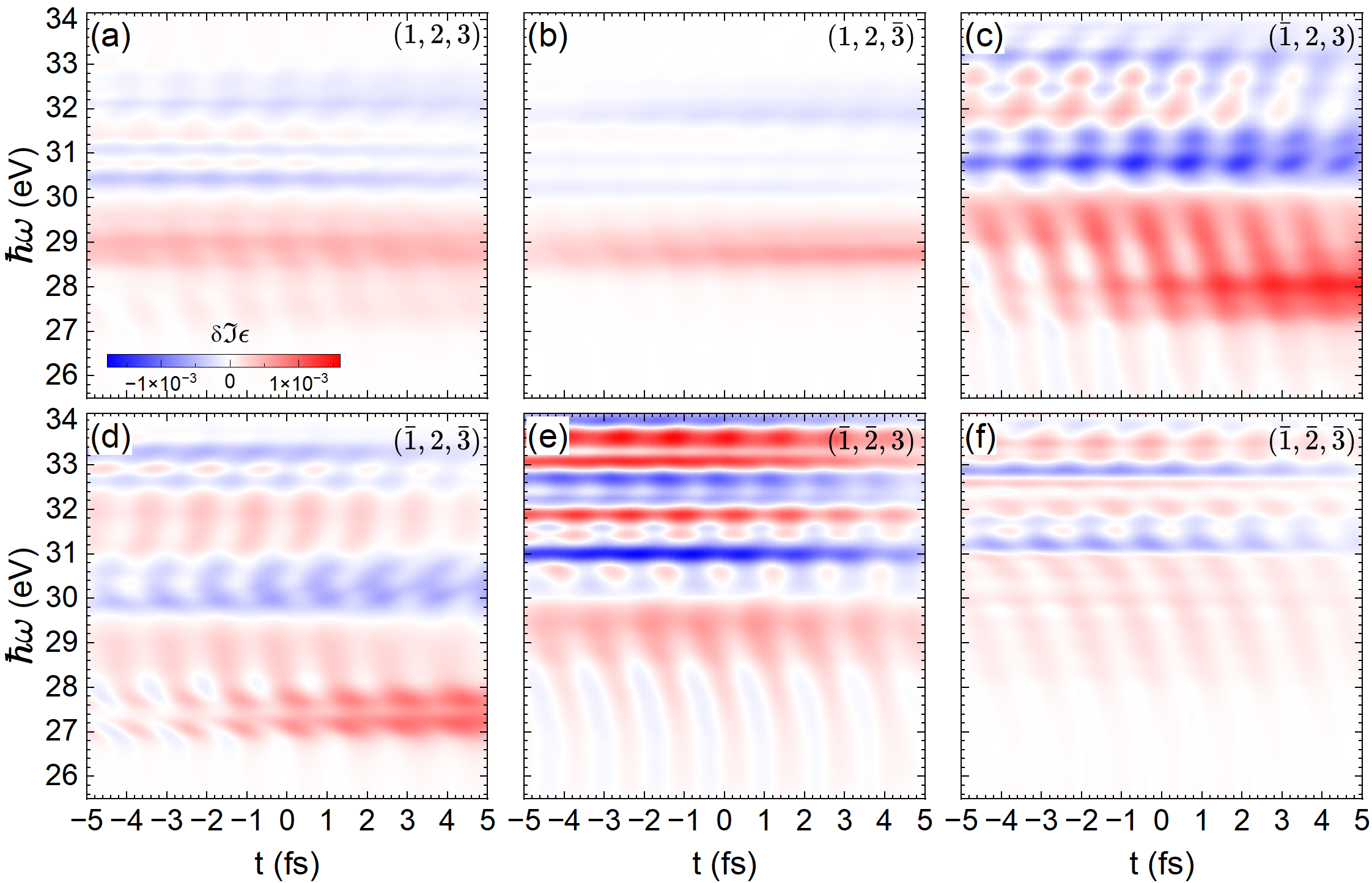}\caption{Decomposition of $\delta\Im\epsilon(\omega,t_{\mathrm{pr}})$ by resonance
class. Each panel corresponds to a class of $\mathbf{k}$-points with
a specific combination of 1-, 2-, and 3-photon resonances, following
Fig.~\ref{fig:res_K}(c) and the notation of Table~\ref{tab:Number-of--points}.
The two classes $(1,\bar{2},3)$, which contain only 16 $\mathbf{k}$-points,
and $(1,\bar{2},\bar{3})$, which contain no $\mathbf{k}$-points,
have negligible and zero contributions to the transient optical features,
respectively, and are therefore not shown. Panel (f) shows the contribution
from the complement of the union of the 1-, 2-, and 3-photon-resonant
sets, comprising both higher-order-resonant and off-resonant $\mathbf{k}$-points.
}\label{fig:Optics_np_k}
\end{figure*}

We now resolve the transient optical response into the complementary
resonance classes introduced above, allowing the contribution of each
momentum-space sector to be examined separately. Figure~\ref{fig:Optics_np_k}
summarizes the resulting decomposition into the mutually exclusive
resonance classes defined in Fig.~\ref{fig:res_K}(c) and Table~\ref{tab:Number-of--points}.
The classes $(1,\bar{2},3)$ and $(1,\bar{2},\bar{3})$ are omitted
because the first contains only 16 sampled points and has a negligible
transient optical contribution, whereas the second is empty.

These resonance classes constitute a complete momentum-space decomposition
of the transient optical response. Their individual analysis therefore
identifies how different resonant regions of the BZ contribute to
the measured response and how the balance between real and virtual
pump-induced contributions evolves across momentum space.

Comparison with the full $\delta\Im\epsilon(\omega,t_{\mathrm{pr}})$
{[}Fig.~\ref{fig:exp_theo_refl}(c){]} reveals how the different
resonance classes contribute to distinct parts of the transient optical
response. Several clear trends emerge. Classes containing 1- and 2-photon
resonances contribute strongly to the structures above the crossover
near $29.8$ eV, whereas subsets selected by higher resonance orders
become increasingly visible at larger probe energies. This correlation
follows from the fact that each pump-resonance class have different
semi-core--to--band transition energies in the probe spectrum. This
energy-dependent hierarchy reflects the selective coupling of different
resonance orders to distinct regions of the electronic structure.
Moreover, these resonance classes form a complete and complementary
decomposition of momentum space, so that summing their contributions
reconstructs the full response.

Panel (e) provides perhaps the clearest evidence for the distinction
between real and virtual pump-induced contributions. Although the
$(\bar{1},\bar{2},3)$-points in this class contribute only marginally
to the residual excitation population $N^{\mathrm{res}}$ {[}Fig.~\ref{fig:res_K}(c)
and Table~\ref{tab:Number-of--points}{]}, they make a non-negligible
contribution to the transient optical response. This observation strongly
supports the interpretation that the optical weight not accounted
for by the real-carrier excitation is predominantly associated with
virtual pump-induced processes. Importantly, these processes remain
concentrated within the same resonant momentum-space regions identified
through the residual populations, rather than being in the off-resonant
regions.

Panel (f) completes the momentum-space decomposition by collecting
all remaining contributions, namely those arising from momentum points
with at least 4-photon resonances together with the off-resonant regions.
Comparison with Fig.~\ref{fig:Optics_res_k}(a) immediately shows
that the higher-order resonant contribution is not negligible. Together
with the analysis of panel (e), this result reinforces the conclusion
that the transient optical weight beyond the real pump-induced carrier
dynamics remains predominantly associated with resonant momentum-space
regions through virtual pump-induced processes.

\begin{figure}
\centering{}\includegraphics[width=8cm]{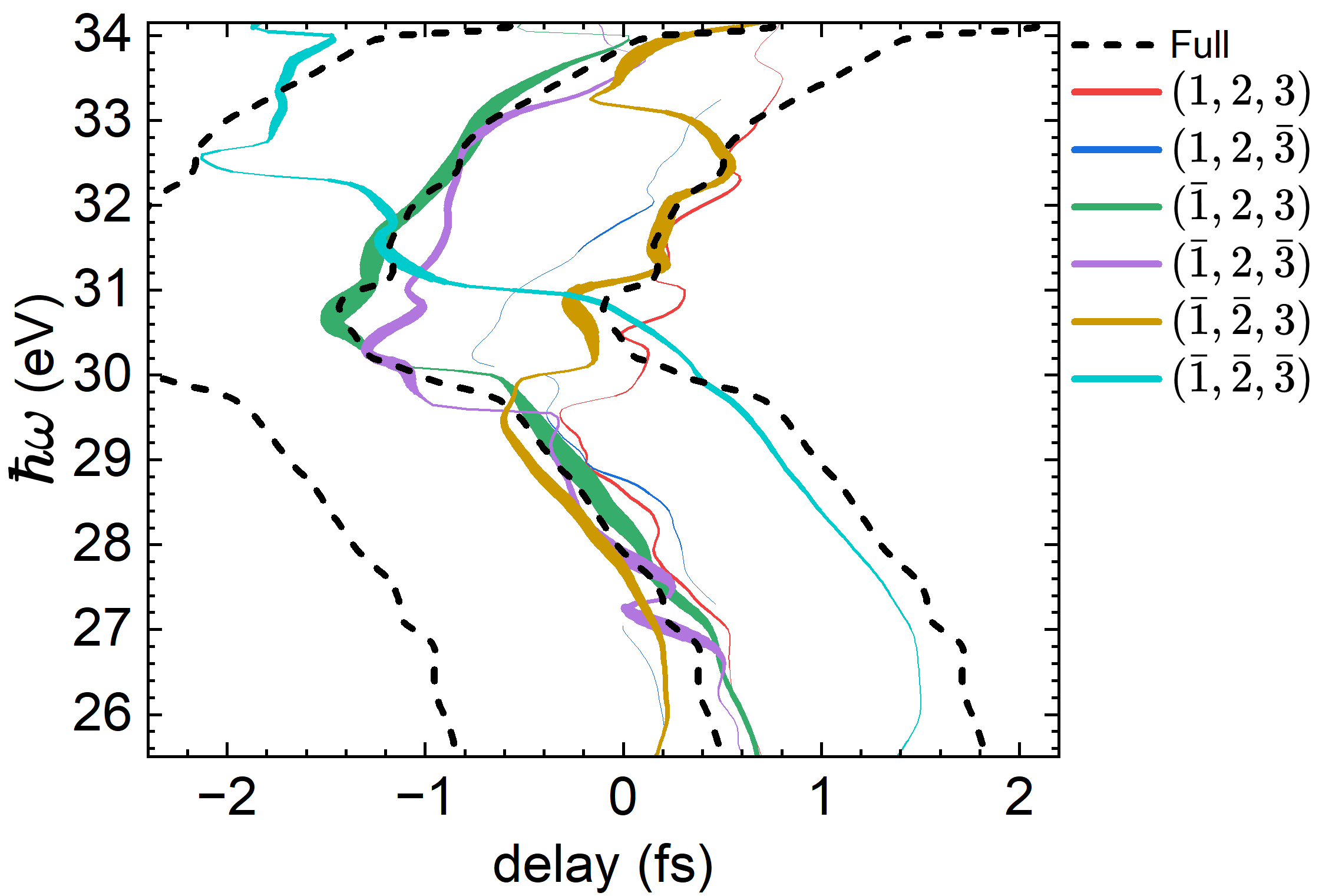}\caption{Time delay of the $2\omega_{\mathrm{pu}}$ component of $\delta\Im\epsilon(\omega,t_{\mathrm{pr}})$
for the different resonance classes shown in Fig.~\ref{fig:Optics_np_k},
compared with the delay extracted from the full response, which is
shown with a periodic repetition every $1.33$ fs. Line thickness
(except for the full response) is proportional to the amplitude of
the $2\omega_{\mathrm{pu}}$ component. }\label{fig:gls}
\end{figure}

The fast oscillations of the optical response of pumped germanium
are dominated by the $2\omega_{\mathrm{pu}}$ component \cite{inzani2023field,inzani2026attosecond}.
To determine its phase relative to the pump-intensity-density envelope,
which is proportional to the squared pump vector potential, $A^{2}_{\mathrm{pu}}(t)$,
we Fourier transform $\delta\Im\epsilon(\omega,t_{\mathrm{pr}})$
with respect to $t_{\mathrm{pr}}$ and extract the phase of the $2\omega_{\mathrm{pu}}$
component, $\phi_{2\omega_{\mathrm{pu}}}$. The corresponding phase-derived
time shift is $\phi_{2\omega_{\mathrm{pu}}}/2\omega_{\mathrm{pu}}$.
For the sinusoidal form used in $A_{\mathrm{pu}}(t)$, applying the
same procedure to $A^{2}_{\mathrm{pu}}(t)$ gives $\pi/2\omega_{\mathrm{pu}}=T_{\mathrm{pu}}/4$,
where $T_{\mathrm{pu}}=2\pi/\omega_{\mathrm{pu}}$. We therefore define
the relative delay as $\phi_{2\omega_{\mathrm{pu}}}/2\omega_{\mathrm{pu}}-T_{\mathrm{pu}}/4$.
Because the phase is defined modulo $2\pi$, this delay is defined
modulo $T_{\mathrm{pu}}/2=1.33$ fs. 

Figure~\ref{fig:gls} compares the resulting class-resolved delays
with the delay extracted from the full response. Periodic replicas
of the latter are shown every $1.33$ fs to display the phase ambiguity.
For each class-resolved curve, the line thickness is proportional
to the absolute amplitude of its $2\omega_{\mathrm{pu}}$ component.

The phase-delay analysis shows that the momentum-space organization
identified above, extends naturally to the ultrafast dynamics. Several
observations support this conclusion. Whenever a class has a sizable
oscillatory amplitude, its delay generally tracks the one of the full
response. Thus, the momentum-space subsets that dominate a given spectral
interval also reproduce the phase of the corresponding sub-cycle oscillation.
Furthermore, the same energy-dependent hierarchy identified in the
transient optical response is also reflected in the phase dynamics:
higher-order resonant $\mathbf{k}$-points give time-delays closer
to the full one at the higher energies, consistently with the spectral
decomposition in Fig.~\ref{fig:Optics_np_k}. Below the crossover
near the CB absorption edge, the class-resolved delays are close to
each other. This behavior is consistent with the interpretation of
Ref.~\cite{inzani2026attosecond} that the corresponding VB-final-state
region is less sensitive to virtual contributions, although similarity
of the delays alone does not constitute an independent decomposition
into real and virtual processes.

\section{Summary and Conclusions}\label{sec:Conclusions}

In this work, we developed and applied a momentum- and $l$-photon-resonance-resolved
framework for identifying the origin of transient optical responses
in driven solids, using pumped germanium as a representative case
study. Using the DPOA together with generalized linear response theory,
we decomposed the differential imaginary part of the dielectric function
into different resonance classes, thereby revealing how different
regions of momentum space contribute to the transient optical response
and to the interplay between real and virtual pump-induced processes.

Taken together, these results establish a coherent picture of how
the resonance-class decomposition organizes the transient optical
response in momentum space. First, we showed that the transient optical
response is governed predominantly by resonant momentum-space regions,
whereas off-resonant regions make only a negligible contribution.
This establishes the resonance-class decomposition as the natural
momentum-space organization of the response. Second, we found that
the real charge excitations alone do not determine the transient optical
response. Although two-photon-resonant momentum points host more than
98\% of the residual excitations (which are resulted from real-charge
excitations), restricting the analysis to those regions is insufficient
to reproduce the complete optical response. Third, we showed that
resonant momentum points with negligible residual excitation populations
can nevertheless contribute significantly to the transient optical
response. This indicates that the optical weight beyond that associated
with the real carrier excitation remains predominantly concentrated
within resonant momentum-space regions and is naturally attributed
to virtual pump-induced processes.

The class decomposition also clarifies the energy dependence of the
response. Different pump-resonance conditions select different regions
of momentum space, and these regions possess different semi-core--to--band
probe-transition energies. Finally, we showed that the same momentum-space
organization also governs the ultrafast phase dynamics. Whenever a
resonance class exhibits a strong $2\omega_{\mathrm{pu}}$ component,
its time delay closely follows that of the full response, indicating
that the dominant oscillatory dynamics originate from the same resonant
momentum-space regions responsible for the transient optical response.

The resonance-resolved construction provides a systematic route for
connecting microscopic populations, coherent dynamics, and optical
observables in realistic band structures. In summary, our results
establish a detailed momentum- and $l$-photon-resonance-resolved
picture of how resonance classes organize the transient optical response
of driven semiconductors, encompassing both real and virtual pump-induced
carrier excitation contributions. Beyond the specific case of germanium,
the resonance-class decomposition introduced here provides a general
framework for identifying the momentum-space origin of transient optical
responses and for disentangling the respective roles of real and virtual
pump-induced processes in real materials.
\begin{acknowledgments}
AE and AA acknowledge support by MUR under Project PNRR MUR Missione
4 (SPOKE 2) TOPQIN ``TOPological Qubit In driveN and reconfigurable
heterostructures''. 
\end{acknowledgments}

 \bibliographystyle{apsrev4-2}
\bibliography{biblio}

\end{document}